\documentstyle[times,pramana,epsfig,floats]{ias}

\def\ds{\displaystyle}
\def\beq{\begin{equation}}
\def\eeq{\end{equation}}
\def\bea{\begin{eqnarray}}
\def\eea{\end{eqnarray}}
\def\beeq{\begin{eqnarray}}
\def\eeeq{\end{eqnarray}}
\def\la{\langle}
\def\ra{\rangle}
\def\gggg{\gamma \gamma \to \gamma \gamma}

\begin{document}

\title{\boldmath Little Higgs model effects in $\gamma \gamma \to \gamma \gamma$}
\author{S. Rai Choudhury \& Ashok Goyal}
\address{Department of Physics \& Astrophysics, University of Delhi, Delhi - 110 007,India}
\author{A. S. Cornell}
\address{Yukawa Institute of Theoretical Physics (YITP), Kyoto, Japan}
\author{Naveen Gaur}
\address{Theory Group, KEK, Tsukuba, Ibaraki 305-0801, Japan}


\abstract{Though the predictions of the Standard Model (SM) are in
  excellent agreement with experiments there are still several
  theoretical problems associated with the Higgs sector of the SM,
  where it is widely believed that some ``{\it new physics}'' will take
  over at the TeV scale. One beyond the SM theory which resolves these
  problems is the Little Higgs (LH) model. In this work we have
  investigated the effects of the LH model on $\gggg$ scattering
  \cite{Choudhury:2006xa}.}    

\maketitle


\section{Introduction}

\par It has been known for some time that the $\gggg$ scattering
amplitude at high energies will be a very useful tool in the search
for new particles and interactions in an $e^+ e^-$ linear collider
operated in the $\gamma \gamma$ mode. In the SM the $\gggg$ amplitudes
will have one-loop contributions mediated by charged fermions (leptons
and quarks) and $W$-bosons. At large energies ($\sqrt{s_{\gamma
    \gamma}} \ge 250$GeV) it is known that the $W$ contributions 
dominate over the fermionic contributions and that the dominant
amplitudes are predominantly imaginary. Therefore we expect that
any new physics effects in the $\gggg$ process may come from the
interference terms between the predominantly imaginary SM amplitudes
and new physics effects to these amplitudes.   

\par The SM has been very successful in explaining all electroweak interactions probed so far, where the SM requires a Higgs scalar field to achieve the electro-weak symmetry breaking. Note that the mass of the Higgs scalar is not protected by any symmetry. In fact the Higgs mass diverges quadratically when quantum corrections in the SM are taken into account. This gives rise to a ``{\sl fine tuning}'' problem in the SM. The precision electroweak data demands the lightest Higgs boson mass be $\sim 200$ GeV! In order for this to happen we need to invoke some symmetries which will protect the Higgs mass to a much higher scale (possibly GUT scale). To resolve the ``{\sl fine tuning}'' problem it is expected that some new physics should takeover from the SM at the TeV scale. The favoured models, {\sl Supersymmetry}, addresses this problem by introducing a symmetry between bosons and fermions. Recently a new approach has been advocated, the approach popularly known as the {\it ``Little Higgs  models''}, which addresses some of the problems in the SM by making the Higgs boson a  pseudo-Goldstone boson of a symmetry which is broken at some higher scale $\Lambda$. For a review of the LH models see references \cite{Schmaltz:2005ky,Hubisz:2005tx}. However, LH models were serverely constrained by precision EW data. The basic problem with these kinds of models was the way in which new physics was coupled to the SM. To resolve these problems a class of models with another symmetry, named {\sl T-parity}, was introduced. These classes of models were investigated in reference \cite{Hubisz:2005tx}. These {\sl T-parity} models had another advantage in that they provided a very useful dark matter candidate. In our work \cite{Choudhury:2006xa} we have analzed $\gggg$ in both the LH and LH with T-parity models.  
%
%

\section{The $\gamma \gamma \to \gamma \gamma$ cross-sections}

\par The process
$
\gamma (p_1, \lambda_1) \gamma (p_2, \lambda_2) \to \gamma (p_3,
\lambda_3) \gamma (p_4, \lambda_4)      
$
can be represented by sixteen possible helicity amplitudes
$F_{\lambda_1 \lambda_2 \lambda_3 \lambda_4}(\hat{s}, \hat{t},
\hat{u})$, where the $p_i$ and $\lambda_i$ represent the respective
momenta and helicities; the $\hat{s}$, $\hat{t}$ and $\hat{u}$ are the
usual Mandelstam variables. By the use of Bose statistics, crossing
symmetries and demanding parity and time-invariance, these sixteen
possible helicity amplitudes can be expressed in terms of just three
amplitudes, namely (the relationships between the various helicity
amplitudes is given in appendix A of our paper
\cite{Choudhury:2006xa}) 
$
F_{++++}(\hat{s}, \hat{t}, \hat{u}) , F_{+++-}(\hat{s},
\hat{t}, \hat{u}) , F_{++--}(\hat{s}, \hat{t}, \hat{u})
$. 
As such, the cross-section for this process can be expressed as
\cite{Gounaris:1999gh} : 
\vskip -.2cm
\bea
\ds \frac{d \sigma}{d\tau d\cos\theta^*} & = & \ds
\frac{d\bar{L}_{\gamma\gamma}}{d\tau} \left\{ \frac{d
\bar{\sigma}_0}{d\cos\theta^*} + \la \xi_2 \xi'_2 \ra \frac{d
\bar{\sigma}_{22}}{d\cos\theta^*} + \left[ \la \xi_3 \ra \cos 2 \phi +
\la \xi'_3 \ra \cos 2 \phi \right] \right. \nonumber \\  
& & \hspace{0.5cm} \times \ds \frac{d \bar{\sigma}_3}{d\cos\theta^*} +
\la \xi_3 \xi'_3 \ra \left[ \frac{d \bar{\sigma}_{33}}{d\cos\theta^*}
\cos 2 ( \phi + \phi' ) + \frac{d \bar{\sigma}'_{33}}{d \cos\theta^*}
\cos 2 ( \phi - \phi' ) \right] \nonumber \\  
& & \left. \hspace{0.5cm} \ds + \left[ \la \xi_2 \xi'_3 \ra \sin 2
\phi' - \la \xi_3 \xi'_2 \ra \sin 2 \phi' \right] \frac{d
\bar{\sigma}_{23}}{d \cos\theta^*} \right\} ,  
\label{one}
\eea
\vskip -.2cm
\noindent where $d \bar{L}_{\gamma \gamma}$ describes the photon-photon
luminosity in the $\gamma \gamma$ mode and $\tau = s_{\gamma
  \gamma}/s_{e e}$. Note that $\xi_2$, $\xi'_2$, $\xi_3$ and $\xi'_3$
are the Stokes parameters. 
To obtain the total cross-section from the above expressions the
integration over $cos\theta^*$ has to be done in the range $0 \le
\cos\theta^* \le 1$. However, the whole range of $\theta^*$ will not
be experimentally observable, hence, for our numerical estimates we
will restrict the scattering angle to $|\cos\theta^*| \le
\sqrt{3}/2$. The process $\gggg$ proceeds through the mediation of
charged particles. In the SM these charged particles were charged
gauge bosons ($W$), quarks and charged leptons. In the LH model, in
addition to the charged gauge bosons and fermions, we also have
charged scalars. The analytical expressions of the contributions from
fermions, gauge bosons and scalars to the helicity amplitudes are
given in reference \cite{Gounaris:1999gh} and are quoted in Appendix A of our paper
\cite{Choudhury:2006xa}. In our work we have analyzed the effects of the
LH models on various polarized cross-sections defined in eqn(\ref{one}) .

%
%
\section{Results and Conclusions}

\par As the $\gggg$ scattering proceeds through loops, both in the SM
and in the LH models (where these loops intermediate particles are
pair produced), in the SM these are dominated by $W$ loops, leading to
a peak in the SM cross-sections around the threshold of the $W$ pair
production \cite{Gounaris:1999gh}. Similarly, in the LH model, the
dominant contribution will come from the new heavy $W$-boson and the
Higgs particles (especially those that are doubly 
charged, as the amplitudes are proportional to the fourth power of the
charge), once we exceed the threshold for the pair production of these
particles. As such, we have plotted the various cross-sections for a
range of energies ($\sqrt{s_{\gamma\gamma}}$) well above the threshold
for the SM $W$-bosons, but in the vicinity of the pair production
energy for the new particles in the LH models. Note further, that we
have integrated our differential cross-sections in the angular range
$30^\circ \le \theta^* \le 150^\circ$.    

\par As expected the deviation in the SM value of the cross-sections
becomes visible around 
the threshold of the pair production of LH particles, where the
present constraints on the LH models forces the masses of all the new
heavy particles to be of the order of TeV. 

\par In all cases we can get substantial deviations in the
cross-sections due to LH effects, however, the $\sigma_3$ and
$\sigma'_{33}$ provide the most interesting results (as given in
figures 1,2), where the $\sigma_3$ is the only cross-section with
pronounced ``{\it dips}''. The location of these ``{\it dips}'' being  
dependent on the model parameters. The other feature of note in these
plots are the pronounced peaks in the $\sigma'_{33}$
cross-section. The SM values of the cross-sections $\sigma_3$ and
$\sigma'_{33}$ are relatively small as compared to the other
cross-sections, however, the LH effects in these two cross-sections
are very striking. These effects mainly depend upon the LH parameter
$f$ (the symmetry breaking scale of the global symmetry). 

\par Though the results we have presented are rather generic and can
be used as a probe for heavy charged gauge bosons and charged
scalars. In our results we have tried to focus ourselves to the range
of cm energy ($\sqrt{s_{\gamma\gamma}}$) which is close to the
threshold of the pair production of the particles. The deviations from
SM results as shown will not be observable in the proposed
International Linear Collider (ILC), but will be easily probed in a
multi-TeV $e^+ e^-$ Compact linear collider (CLIC); where it is
proposed to build an $e^+ e^-$ linear collider with a center of mass
energy from 0.5 - 3TeV. Generically such a mode should lead to $\gamma
\gamma$ collisions at cm energies $E^{\gamma\gamma}_{cm} \le 0.8
E^{ee}_{cm}$. Furthermore, the polarized cross-sections $\sigma_3$ and
$\sigma'_{33}$ can be used to test the spin structure of the particle
loops which are responsible for the $\gggg$ process
\cite{Gounaris:1999gh}. In summary the $\gggg$ process is a very clean
process which shall provide a very useful tool for testing LH type
models. 

%
%

\begin{figure}[h,t,b]
\hskip -1cm
\epsfig{file=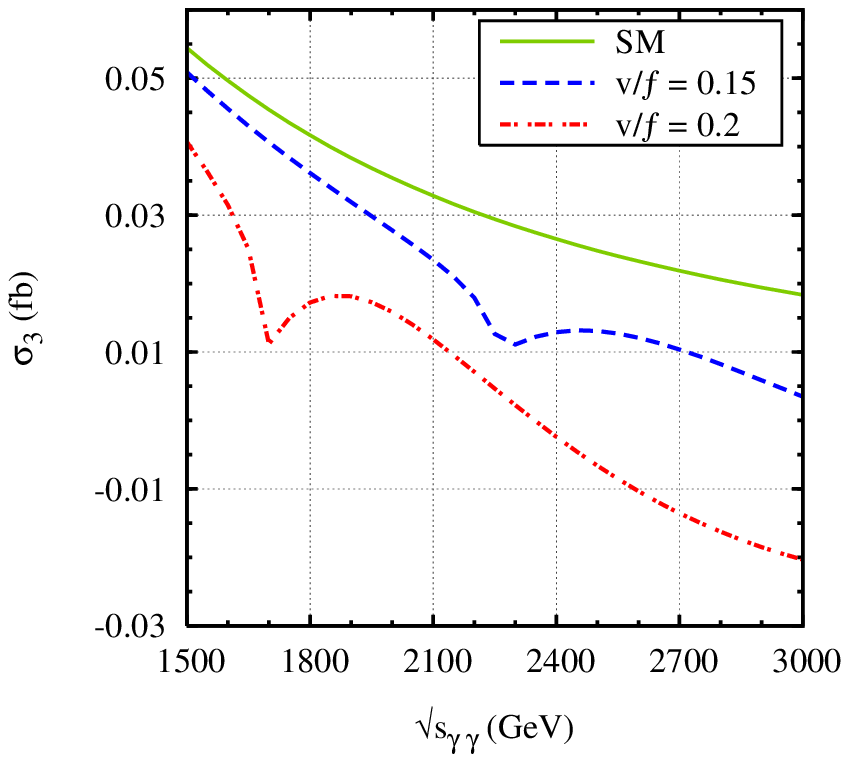,width=.6\textwidth} \hskip -2cm
\epsfig{file=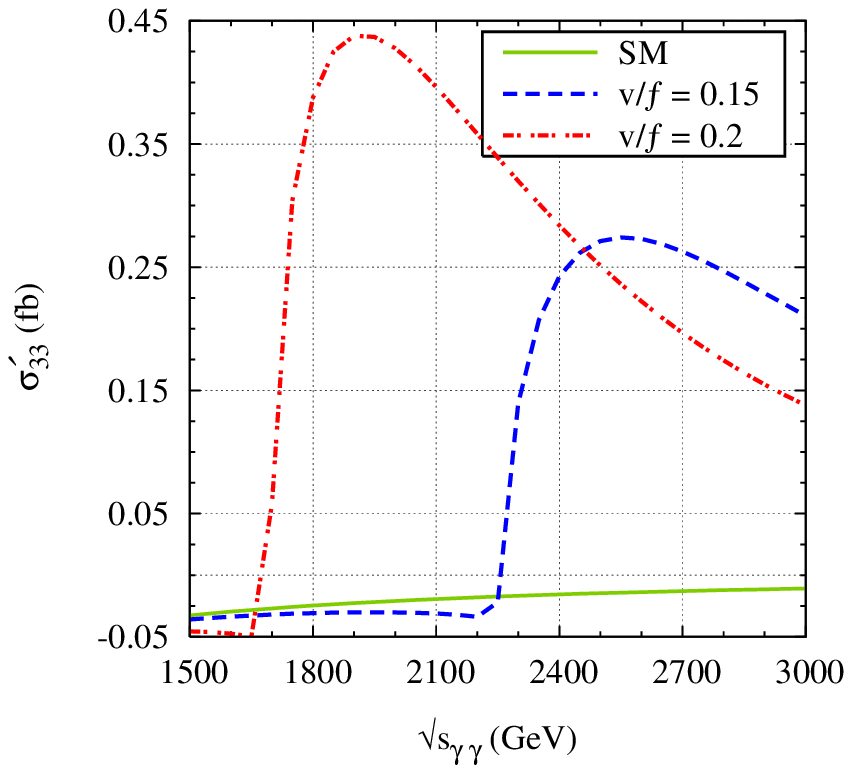,width=.6\textwidth} \hskip -1cm
\caption{\it Results for the cross-sections integrated in the range
$30 \le \theta^* \le 150$ for various values of $v/f$. Other LH model
parameters are: $x_L = 0.2, s = s' = 0.6$.}  
\label{fig:lh_intcross_1}
\end{figure}

\begin{figure}[h]
\hskip -1cm
\epsfig{file=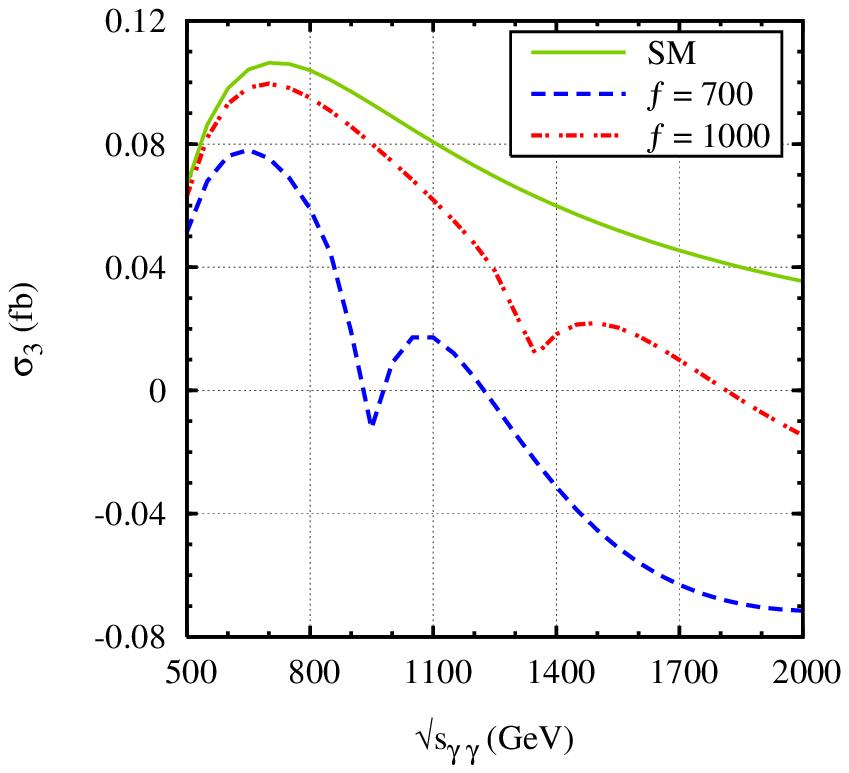,width=.6\textwidth} \hskip -2cm
\epsfig{file=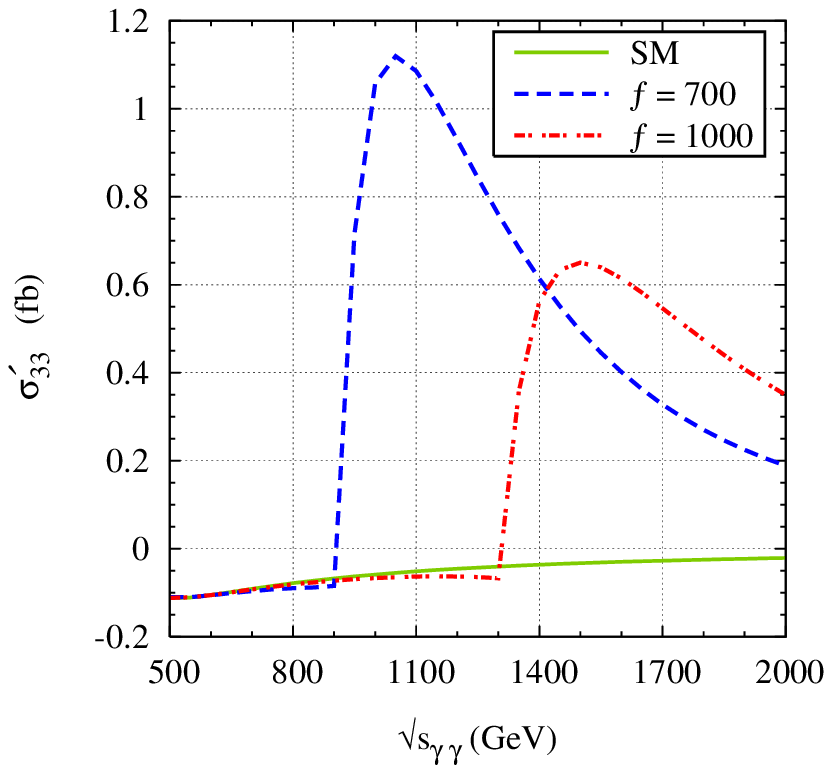,width=.6\textwidth} \hskip -1cm
\caption{\it Results for the cross-sections integrated in the range
$30 \le \theta^* \le 150$ for various values of $v/f$ in LH model 
with T-parity.}
\label{fig:lh_intcross_2}
\end{figure}

%
%

\section*{Acknowledgements}
The work of SRC, NG and AKG was supported by the Department of Science
\& Technology (DST), India under grant no SP/S2/K-20/99. The work of
ASC was supported by the Japan Society for the Promotion of Science
(JSPS), under fellowship no P04764. The work of NG was supported by JSPS, 
under fellowship no. P06043.

%
%

\end{document}